# Weak Localization and Electron-electron Interactions

# in Few Layer Black Phosphorus Devices

Yanmeng Shi[1][§], Nathaniel Gillgren[1][§], Timothy Espiritu[1], Son Tran[1], Jiawei Yang[1], Kenji Watanabe[2], Takahashi Taniguchi[2], Chun Ning Lau[1][*]

[1] Department of Physics and Astronomy, University of California, Riverside, Riverside, CA 92521

[2] National Institute for Materials Science, 1-1 Namiki Tsukuba, Ibaraki 305-0044, Japan.

[§]These authors contributed equally to this work. The order was decided by a coin toss.

ABSTRACT

Few layer phosphorene(FLP) devices are extensively studied due to its unique electronic properties and potential applications on nano-electronics . Here we present magnetotransport studies which reveal electron-electron interactions as the dominant scattering mechanism in hexagonal boron nitride-encapsulated FLP devices. From weak localization measurements, we estimate the electron dephasing length to be 30 to 100 nm at low temperatures, which exhibits a strong dependence on carrier density *n* and a power-law dependence on temperature ($\sim T^{-0.4}$). These results establish that the dominant scattering mechanism in FLP is electron-electron interactions.

The experimental isolation of graphene on insulating substrates[1] over a decade ago has led to the explosively increasing interest in two-dimensional (2D) materials[2]. A plethora of 2D atomic layers have been investigated, with properties ranging from semi-metallic, metallic, semi-conducting, insulating and superconducting, with many fascinating electronic, optical, thermal and mechanical properties.

One of the recent additions to the family of 2D materials is phosphorene, which is single- or few- atomic layers of black phosphorus (BP). BP is the most stable form of elemental phosphorus, consisting of layers held together by weak van der Waal's forces. Within each layer, the phosphorus atoms are arranged in a puckered structure[3, 4] It has recently piqued the interest of the scientific community due to its high mobility[5], direct band gap that is tunable by thickness or strain[6-12], and large in-plane anisotropy[4, 7, 12, 13]. These properties make BP a highly attractive candidate for electronics, thermal and optoelectronics applications, as well as a

[*] Email: lau@physics.ucr.edu

model system for interesting physics such as anisotropic quantum Hall effect[14] and emergence of topological order under an electric field[15, 16]

The first generation of few-layer phosphorene (FLP) devices were fabricated on Si/SiO$_2$ substrates, with reported hole mobility ~ 300 – 1000 cm$^2$V$^{-1}$s$^{-1}$. Recent experiments[14, 17] have pushed the charge carrier mobility to ~ 4000-6000 cm$^2$V$^{-1}$s$^{-1}$ at low temperatures by using hexagonal boron nitride (hBN) as substrates, enabling observation of Shubnikov de Haas oscillations[17-19], and more recently quantum hall effect[14]. However, despite the recent progress, the mobility of FLP is still an order of magnitude smaller than that in bulk, ~ 50-60,000 cm$^2$V$^{-1}$s$^{-1}$.[1] In particular, scattering mechanisms in FLP are not well-understood. In this article, we use weak localization to explore the inelastic scattering mechanism in hBN-encapsulated FLP at low temperature. The phase coherence length $L_\varphi$ of charges is measured to be 30 – 100 nm at 1.5 K, and proportional to $T^{-1/2}$. As charge density $n$ increases, $L_\varphi$ exhibits a superlinear dependence on $n$. These observations are consistent with the theory of electron-electron interactions with small momentum transfer, which gives rise to electronic dephasing in FLP at low temperatures.

To fabricate hBN-encapsulated FLP devices[17, 20], we exfoliate FLP sheets from bulk crystals inside a VTI glove box onto PDMS stamps, and thin hBN layers on Si substrates with 300 nm of SiO$_2$. Using a home-built micromanipulator stage, we align FLP and hBN sheets and bring them into contact. The PDMS stamp is then peeled off, leaving the FLP/hBN stack on Si/SiO$_2$ substrate, and the transfer procedure is repeated to place a few-layer hBN sheet onto the stack. The entire assembly process is performed inside the glovebox. The completed hBN/FLP/hBN heterostructure is patterned into Hall-bar geometry by standard electron beam lithography, and the top BN layer is etched by SF$_6$ gas in an ICP etcher to expose the FLP sheets. To minimize degradation, Cr/Au contacts are deposited immediately after the etching process. The optical image of a typical device is shown in Figure 1a. The devices are measured in He$^3$ refrigerator or in a pumped He$^4$ cryostat with a variable temperature insert. Here we present transport data from two different devices that are ~ 20 nm thick, with mobility up to 1700 cm$^2$/Vs.

FLP has a thickness-dependent band gap[11]. For FLP that are more than 5 layers, the gap is similar to that of bulk, ~ 0.3 eV[11]. Fig. 1b displays the two-terminal conductivity $\sigma$ of device A as a function of applied gate voltage ($V_g$) at $T$=300mK, and Fig. 1c displays the four-terminal conductivity of device B at $T$=1.5K. $\sigma$~0 for both devices, when the Fermi level is within the band gap. For $V_g$<-15V, $\sigma$ increases linearly (device A) or superlinearly (device *B*) with $V_g$, indicating hole mobility of ~ 500 cm$^2$/Vs and 1700 cm$^2$/Vs, respectively. We note that the apparent doping of the device, *i.e.* the offset of the charge neutrality points from zero gate voltage, may result from the charge transfer from PDMS during the fabrication process, or presence of small amount of impurities on the hBN sheets. In both devices, conductivity of electron-doped regime is significantly lower, which is likely due to the formation of Schottky barriers at electrode-BP interfaces. Hence we focus on transport properties in the *p*-doped regime.

In the highly *p*-doped regime, device conductivity significantly exceeds $\sigma_q$, where $\sigma_q=e^2/h\sim39$ µS is the conductance quantum, thus the device is in the metallic regime (here *h* is Planck's constant, *e* the electron charge). However, $\sigma$ decreases slightly with *T*. This is a signature of electron interactions in disordered 2D thin films. In fact, $\sigma$ is expected to exhibit a logarithmic dependence on *T*,

$$\sigma = \sigma_0 + C\frac{2e^2}{\pi h}\ln\left(\frac{T}{T_0}\right) \qquad (1)$$

where $\sigma_0$ is the "intrinsic" metallic conductivity, $T_0$ is a characteristic temperature estimated to be $\hbar/k_B\tau_0$, $k_B$ the Boltzmann's constant, $\tau_0$ the electron scattering time, and *C* a dimensionless constant that is of order unity depending on the scattering mechanism[21]. This logarithmic dependence is borne out by experimental data from device B, shown as circles in Fig. 1d. At charge density $n\sim -1.75\times10^{16}$ m$^{-2}$, $\sigma_0\sim 0.4$ mS, and using effective mass $m^*\sim 0.26\ m_e$[17], $\tau_0$ is estimated from Drude model to be ~0.21 ps, yielding $T_0 \sim 36$ K. The dashed line is a fit to Eq. (1) with $C\sim1.6$ as the fitting parameter, in agreement with theory. (Device A was only measured at *T*=0.3K and 1.5K, where conductivity remains intrinsic and constant).

To further explore the inelastic scattering mechanism, we employ weak localization measurements by applying a perpendicular magnetic field *B*. Weak localization (WL) is the quantum correction to the classical conductivity of a diffusive system[22, 23]. In a 2D system, due to multiple inelastic scatterings, electrons in a closed trajectory interfere constructively with the time-reversed path, resulting in enhanced backscattering and hence lower conductivity. Application of a small *B* destroys the interference, thus conductivity increases. WL has been widely applied to 2D systems for measuring the inelastic scattering time, characterized by the dephasing time $\tau_\varphi$. When the elastic scattering time is much shorter than the inelastic scattering time, the change in magnetoconductance induced by *B* is given by[23, 24]

$$\Delta\sigma = \sigma(B) - \sigma(B=0) = -\frac{e^2}{\pi h}\left[\ln\left(\frac{B_\varphi}{B}\right) - \Psi\left(\frac{1}{2}+\frac{B_\varphi}{B}\right)\right] \qquad (2)$$

where $\sigma$ is the device conductivity, *h* is Planck's constant, *e* is the electron charge, $\Psi$ is the digamma function, $B_\varphi = \frac{\hbar}{4eL_\varphi^2}$ is the magnetic field required to destroy phase coherence, $L_\varphi=\sqrt{D\tau_\varphi}$, and *D* is the diffusion coefficient.

Figure 2a displays the normalized conductivity $\Delta\sigma$ in units of $\frac{e^2}{\pi h}$ (color) as a function of *n* (vertical axis) and *B* (horizontal axis). As *B* is swept from -0.8T to +0.8T, $\Delta\sigma$ displays positive magnetoconductivity with a minimum at *B*=0T, consistent with the time reversal symmetry breaking of phase coherent back scattering. The magnitude of $\Delta\sigma$ is relatively large (>1) when the device is highly doped, and small (<0.1) when the Fermi level is close to the band edge. Representative line traces $\Delta\sigma(n)$ are shown as solid lines in Fig. 2b, and the dashed lines are fits using Eq. (2). Satisfactory agreement between the data and Eq. (2) are obtained. From the fitting parameter $B_\varphi$, we obtain $L_\varphi \sim 75$ nm at $n=-1.8 \times 10^{12}$ cm$^{-2}$. Using $D=\frac{\hbar}{4m^*}\frac{\sigma}{\sigma_q} \sim 4.3\times10^{-4}$ m$^2$/s, the inelastic scattering time is estimated to be $\tau_\varphi \sim 13$ ps, which is two orders of magnitude longer than the elastic scattering time $\tau_0$. This is consistent with the applicability condition of Eq. (2), and establishes that charge transport in these FLP devices is diffusive but phase coherent over tens of nanometers. Similar calculations yield that, at $n=-1.0$ and $-0.7\times10^{12}$cm$^{-2}$, $L_\varphi \sim 60$ nm and 40 nm and $\tau_\varphi \sim 8.4$ps and 3.7ps, respectively.

To explore the scattering mechanism, we examine the temperature dependence of $L_\varphi$ in device B. In general, the dephasing time $\tau_\varphi \sim T^{-\alpha}$, where the exponent $\alpha$ depends on the scattering mechanisms. In particular, $\alpha \sim 2$ for electron-phonon scattering. On the other hand, if electron-electron interaction is the dominant mechanism, two separate processes may occur depending on the impurity density of the system[23, 25]-- the first involves direct scattering between electrons and large momentum transfer, with a rate that scales with $(k_BT)^2$, so $\alpha=2$; the second process involves small momentum transfer, and considers not individual collision events but instead the interaction of an electron with the fluctuating electromagnetic environment produced by the movement of other electrons. The latter process is similar to that in the Nyquist noise, with a rate that scales linearly with $k_BT$ in 2D, hence $\alpha=1$[23]..

Figure 3 displays $L_\varphi(T)$ at four different hole densities. As expected, $L_\varphi$ increases as $T$ decreases from 15K to 2K, then saturates for $T<2$K. We fit the data points above 2K to a power law dependence $L_\varphi \sim T^{-\beta}$. The measured values of $\beta=\alpha/2$ are found to be $\sim 0.4 \pm 0.02$ for all densities, which is close to the value of $\beta=0.5$ or $\alpha=1$ expected from the theory of electronic interactions with small momentum transfer. Quantitatively, the Altshuler-Aronov-Khmelnitsky theory predicts $\frac{\hbar}{\tau_\varphi} = \frac{k_BT}{\sigma/\sigma_q}\ln(\sigma/\sigma_q)$ [23]. Combined with the expression for $D$, the dephasing length is given by

$$L_\varphi = \frac{\hbar\sigma}{\sigma_q}\left[\ln(\sigma/\sigma_q)4m^*k_BT\right]^{-1/2} \qquad (3)$$

From Eq. (1), $\sigma$ exhibits a weak logarithmic dependence on $T$, thus $L_\varphi$ should scale with $T^{1/2}\ln(T)$, and the $\ln(T)$ term accounts for the observed deviation of $\beta$ from the expected value of 0.5. Thus the $L_\varphi(T)$ data establish that the main dephasing mechanism at low temperatures arise from electron-electron interactions with small momentum transfer, though the saturation of $L_\varphi$ at $T<2$K may suggest a different mechanism at ultra-low temperatures.

Lastly, we explore the dependence of $L_\varphi$ on charge density. Fig. 4a plots $L_\varphi(n)$ for device A at $T$=0.3K, and that for device B at $T$=1.5K and 4K. Clearly, $L_\varphi$ is strongly dependent on carrier density, varying by almost 1 order of magnitude from 30 nm to 110 nm when $n$ increases from 0.5 to $2\times10^{12}$ cm$^{-2}$. In fact, $L_\varphi$ appears to have power-law dependence on $n$, $L_\varphi \sim n^p$, where $p$ appears to be less than 1 for device A at $T$=300 mK, and ~1.5 and 2 for device B at $T$=1.5K and 4K, respectively.

Such $L_\varphi(n)$ dependence can be readily understood from Eq. (3), which shows that $L_\varphi$ should be slightly superlinear in $\sigma$, and the latter is in turn linear or superlinear in $n$. This is explicitly verified by plotting $L_\varphi/\sigma$ for device A at $T$=0.3 K and device B at $T$=1.5K and 4K, respectively (Fig. 4b). All 3 curves are relatively independent of $n$ at relatively high hole density. The slight rises in the curves at lower carrier density is attributed to the larger Schottky barriers and increasing contact resistance towards the band edge. This deviation is largest in the two-terminal data of Device A, and much smaller but still present in the *invasive* four-terminal data of Device B. Taken together, these results again confirm that inelastic scattering processes at low temperature in FLP are dominated by electron-electron interactions

In short, we have observed the weak localization in hBN-encapsulated FLP devices. The dephasing length is measured to be ~30 to 100 nm, and exhibits power-law dependences on temperature and charge density. Our results demonstrate that the main dephasing mechanism in these few layer BP devices is electron-electron interactions. Finally, during the preparation of the manuscript, we become aware of a similar work[26] on thinner (4.5 – 8.2 nm) FLP devices on Si/SiO$_2$ substrates. The similarities between the results from devices on different substrates suggest that, at low temperatures, scattering in these systems with mobility 500-2000 cm$^2$/Vs is dominated by disorder-mediated electron-electron interactions, and not limited by substrates or dependent on thickness (down to ~5 nm). On the other hand, for very thin FLP sheets with significantly larger band gaps and different mobility bottlenecks, other factors such as screening, impurities and electron interactions with large momentum transfer may play different roles. Thus further studies are warranted to reveal scattering mechanism in higher mobility samples or very thin FLP (down to monolayer) devices. Another interesting topic is the angular dependence of scattering and dephasing mechanism, which, given the large lattice anisotropy, may vary along different crystallographic directions and await further investigation.

This work is supported by FAME center, one of six centers of STARnet, a Semiconductor Research Corporation program sponsored by MARCO and DARPA, and by NSF/ECCS 1509958. S.T. is supported by ONR. Growth of hexagonal boron nitride crystals was supported by the Elemental Strategy Initiative conducted by the MEXT, Japan and a Grant-in-Aid for Scientific Research on Innovative Areas "Science of Atomic Layers" from JSPS.

Note added during proof: After acceptance of this manuscript, we became aware of a similar work on weak localization and dephasing in black phosphorus[27].

# References

[1] Novoselov K S, Geim A K, Morozov S V, Jiang D, Zhang Y, Dubonos S V, Grigorieva I V and Firsov A A 2004 Electric field effect in atomically thin carbon films *Science* **306** 666-9
[2] Geim A K and Grigorieva I V 2013 Van der Waals heterostructures *Nature* **499** 419-25
[3] Li L, Yu Y, Ye G J, Ge Q, Ou X, Wu H, Feng D, Chen X H and Zhang Y 2014 Black phosphorus field-effect transistors *Nat Nano* **9** 372-7
[4] Ling X, Wang H, Huang S, Xia F and Dresselhaus M S 2015 The renaissance of black phosphorus *Proceedings of the National Academy of Sciences* **112** 4523-30
[5] Akahama Y, Endo S and Narita S-i 1983 Electrical Properties of Black Phosphorus Single Crystals *Journal of the Physical Society of Japan* **52** 2148-55
[6] Maruyama Y, Suzuki S, Kobayashi K and Tanuma S 1981 Synthesis and some properties of black phosphorus single crystals *Physica B+C* **105** 99-102
[7] Liu H, Neal A T, Zhu Z, Luo Z, Xu X, Tománek D and Ye P D 2014 Phosphorene: An Unexplored 2D Semiconductor with a High Hole Mobility *ACS Nano* **8** 4033-41
[8] Du Y, Ouyang C, Shi S and Lei M 2010 Ab initio studies on atomic and electronic structures of black phosphorus *Journal of Applied Physics* **107** 093718
[9] Keyes R W 1953 The Electrical Properties of Black Phosphorus *Physical Review* **92** 580-4
[10] Warschauer D 1963 Electrical and Optical Properties of Crystalline Black Phosphorus *Journal of Applied Physics* **34** 1853
[11] Tran V, Soklaski R, Liang Y and Yang L 2014 Layer-controlled band gap and anisotropic excitons in few-layer black phosphorus *Physical Review B* **89** 235319
[12] Qiao J, Kong X, Hu Z-X, Yang F and Ji W 2014 High-mobility transport anisotropy and linear dichroism in few-layer black phosphorus. *Nature communications* **5** 4475
[13] Xia F, Wang H and Jia Y 2014 Rediscovering black phosphorus as an anisotropic layered material for optoelectronics and electronics *Nat Commun* **5**
[14] Li L, Yang F, Ye G J, Zhang Z, Zhu Z, Lou W, Zhou X, Li L, Watanabe K, Taniguchi T, Chang K, Wang Y, Chen X H and Zhang Y 2016 Quantum Hall effect in black phosphorus two-dimensional electron system *Nat Nano* **advance online publication**
[15] Liu Q, Zhang X, Abdalla L B, Fazzio A and Zunger A 2015 Switching a normal insulator into a topological insulator via electric field with application to phosphorene. *Nano letters* **15** 1222-8
[16] Kim J, Baik S S, Ryu S H, Sohn Y, Park S, Park B-G, Denlinger J, Yi Y, Choi H J and Kim K S 2015 2D MATERIALS. Observation of tunable band gap and anisotropic Dirac semimetal state in black phosphorus. *Science (New York, N.Y.)* **349** 723-6
[17] Gillgren N, Wickramaratne D, Shi Y, Espiritu T, Yang J, Hu J, Wei J, Liu X, Mao Z, Watanabe K, Taniguchi T, Bockrath M, Barlas Y, Lake R K and Ning Lau C 2014 Gate tunable quantum oscillations in air-stable and high mobility few-layer phosphorene heterostructures *2D Materials* **2** 011001
[18] Li L, Ye G J, Tran V, Fei R, Chen G, Wang H, Wang J, Watanabe K, Taniguchi T, Yang L, Chen X H and Zhang Y 2015 Quantum oscillations in a two-dimensional electron gas in black phosphorus thin films. *Nature nanotechnology* **10** 608-13

[19]  Tayari V, Hemsworth N, Fakih I, Favron A, Gaufrès E, Gervais G, Martel R and Szkopek T 2015 Two-dimensional magnetotransport in a black phosphorus naked quantum well. *Nature communications* **6** 7702
[20]  Castellanos-Gomez A, Buscema M, Molenaar R, Singh V, Janssen L, van der Zant H S J and Steele G A 2014 Deterministic transfer of two-dimensional materials by all-dry viscoelastic stamping *2D Materials* **1** 011002
[21]  Ando T, Fowler A B and Stern F 1982 Electronic properties of two-dimensional systems *Reviews of Modern Physics* **54** 437-672
[22]  Hikami S, Larkin A I and Nagaoka Y 1980 Spin-Orbit Interaction and Magnetoresistance in the Two Dimensional Random System *Progress of Theoretical Physics* **63** 707-10
[23]  Altshuler B L, Aronov a G and Khmelnitsky D E 1982 Effects of electron-electron collisions with small energy transfers on quantum localisation *Journal of Physics C: Solid State Physics* **15** 7367-86
[24]  Bergmann G 1986 Weak Localization in Thin Films *Physica Scripta* **1986** 99
[25]  Fukuyama H and Abrahams E 1983 Inelastic scattering time in two-dimensional disordered metals *Physical Review B* **27** 5976-80
[26]  Yuchen D, Adam T N, Hong Z and Peide D Y 2016 Weak localization in few-layer black phosphorus *2D Materials* **3** 024003
[27]  Hemsworth N, Tayari V, Telesio F, Xiang S, Roddaro S, Caporali M, Ienco A, Serrano-Ruiz M, Peruzzini M, Gervais G, Szkopek T, Heun S, 2016, Dephasing in strongly anisotropic black phosphorus, arXiv:1607.08677

FIG. 1 (a) An optical image of a typical hBN-encapsulated FLG device with hall bar geometry and a top gate. Scale bar: 10 μm. (b) Two-terminal conductivity as a function of back gate voltage of device A at $T = 0.3$K. (c) Four-terminal conductivity as function of back gate voltage of device B at $T = 1.5$K. (d) The conductivity of device B as function of temperature taken at $V_{bg} = -35$V. The dashed line is a fit to Eq.(1).

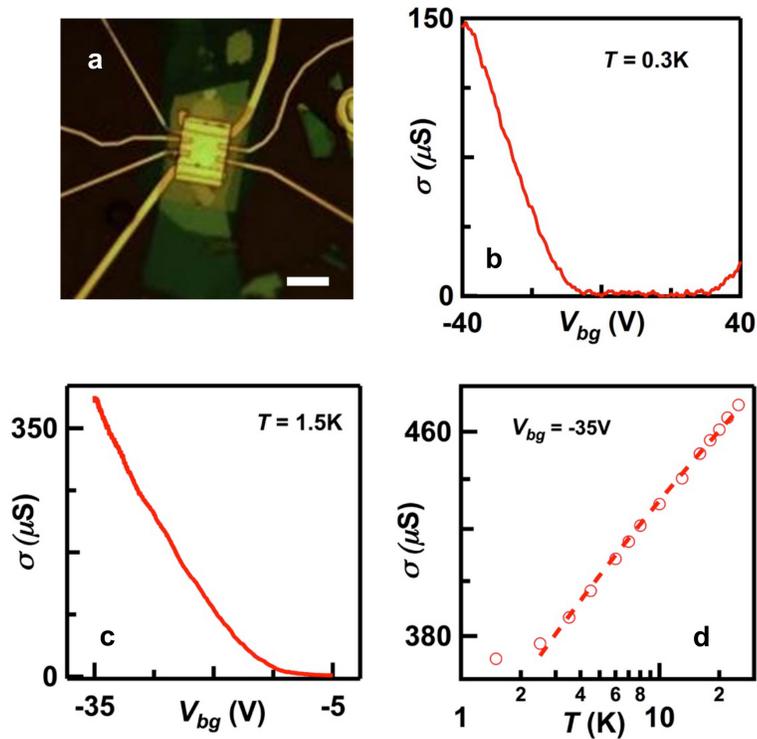

FIG. 2 Weak localization data from device A. (a). Normalized conductivity $\Delta\sigma$ in units of $e^2/\pi h$ vs. carrier density and magnetic field. Note that below $-0.38 \times 10^{-12}$ cm$^{-2}$ the device no longer displays the suppression of weak localization. This is attributed to the device entering the insulating state. (b) Solid lines: line traces $\Delta\sigma(B)$ at -1.8, -1, and -0.7 x0$^{-12}$cm$^{-2}$, respectively (top to bottom). Dotted lines: fits to the data using Eq.(2).

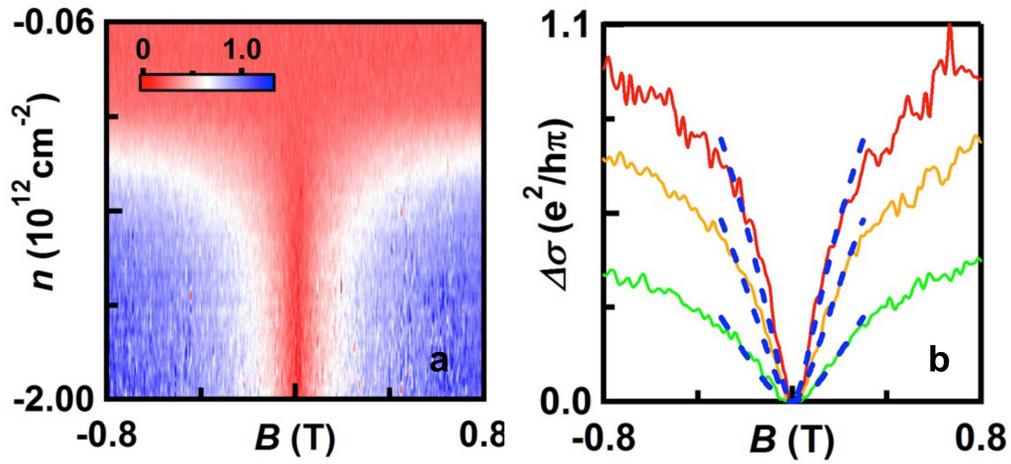

FIG. 3 Temperature dependence of dephasing length of device B at different hole densities. Dashed lines are fits to power-law dependence $T^{-0.4}$

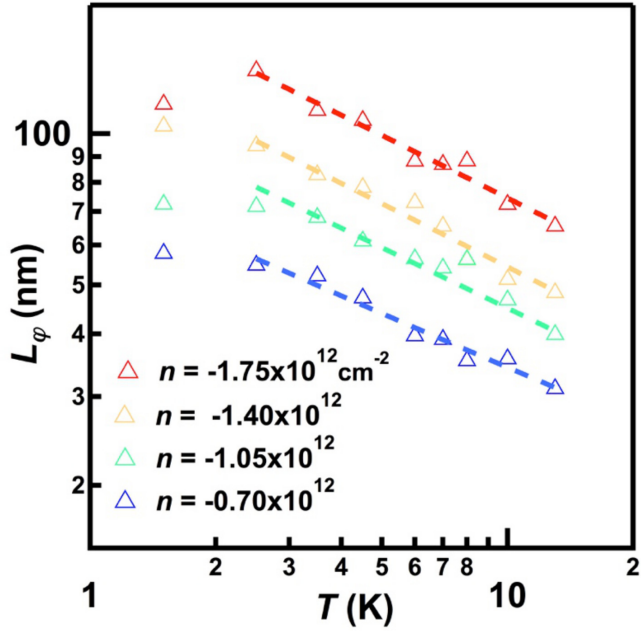

FIG. 4 (a) The dependence of dephasing length $L_\varphi$ on carrier density from Device A at 0.3K(blue), and Device B at 1.5K (red) and 4K (green) respectively. (b) $L_\varphi/\sigma$ (n) for the devices, where $\sigma$ is taken from $B=0$ measurements.

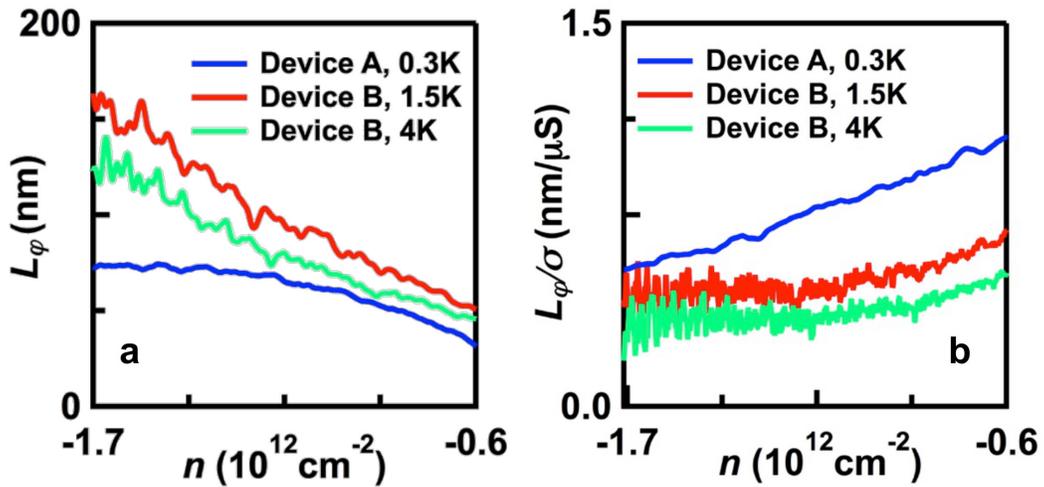